\documentclass[acmlarge]{acmart}
\AtBeginDocument{%
  }

\usepackage{adjustbox}
\usepackage{longtable} 

\begin{document}

\title{From Code to Compliance: Assessing ChatGPT’s Utility in Designing an Accessible Webpage—A Case Study}

\author{Ammar Ahmed}
\affiliation{%
  \institution{Department of Computer Science, Norwegian University of Science
and Technology, Gjøvik}
\city{}
  \country{Norway}
}
\email{ammaa@stud.ntnu.no}
\orcid{0009-0003-9984-4819}

\author{Margarida Fresco}
\affiliation{%
  \institution{Department of Design, Norwegian University of Science
and Technology, Gjøvik}
  \city{}
  \country{Norway}
}
\email{margario@stud.ntnu.no}

\author{Fredrik Forsberg}
\affiliation{%
  \institution{Department of Design, Norwegian University of Science
and Technology, Gjøvik}
\city{}
  \country{Norway}
}
\email{fredrfor@stud.ntnu.no}

\author{Hallvard Grotli}
\affiliation{%
  \institution{Department of Design, Norwegian University of Science
and Technology, Gjøvik}
\city{}
  \country{Norway}
}
\email{grotli321@gmail.com}


\begin{abstract}


 Web accessibility ensures that individuals with disabilities can access and interact with digital content without barriers, yet a significant majority of most used websites fail to meet accessibility standards. This study evaluates ChatGPT's (GPT-4o) ability to generate and improve web pages in line with Web Content Accessibility Guidelines (WCAG). While ChatGPT can effectively address accessibility issues when prompted, its default code often lacks compliance, reflecting limitations in its training data and prevailing inaccessible web practices.
 Automated and manual testing revealed strengths in resolving simple issues but challenges with complex tasks, requiring human oversight and additional iterations. Unlike prior studies, we incorporate manual evaluation, dynamic elements, and use the visual reasoning capability of ChatGPT along with the prompts to fix accessibility issues. Providing screenshots alongside prompts enhances the LLM’s ability to address accessibility issues by allowing it to analyze surrounding components, such as determining appropriate contrast colors. We found that effective prompt engineering, such as providing concise, structured feedback and incorporating visual aids, significantly enhances ChatGPT's performance. These findings highlight the potential and limitations of large language models for accessible web development, offering practical guidance for developers to create more inclusive websites.
\end{abstract}



\maketitle

\section{Introduction}
The goal of web accessibility is to remove barriers that might prevent people with disabilities from accessing or having a full user experience. Effective accessibility integration helps to fulfill the principles of universal design, which is the design of products and environments to be usable by all people, to the greatest extent possible \cite{Mitrasinovic2008}. Instead of addressing accessibility as an afterthought in completed projects, universal accessibility must take a proactive approach from the very beginning of the design and development processes. To address these concerns, the Web Content Accessibility Guidelines (WCAG) were established as an international standard, providing detailed guidance on making web content more accessible \cite{WCAG22}.


Despite the critical importance of accessibility, WebAIM’s 2024 report on one million frequently used homepages indicates that 96\% of these websites were deemed inaccessible \cite{WebAIM2024}. The analysis identified over 56 million accessibility errors, equating to an average of 56.8 errors per page. This highlights a significant issue, particularly given that approximately 16\% of the global population—around 1.3 billion people—live with some form of disability \cite{WHO2023}. Moreover, they identified that the complexity of the analyzed homepages had increased from last year up to 11.8\%. Therefore, addressing these barriers requires deliberate efforts, but the growing complexity of web development has made achieving accessibility a persistent challenge.

In recent years, generative AI has made significant strides, not only in producing synthetic videos and text but also recently in generating accurate code \cite{Cooper2024, Jiang2024, Dou2024}, with many developers incorporating the use of AI in several software development stages \cite{Sergeyuk2024}. The field is rapidly evolving with new findings published every month \cite{Zhao2023_survey, Zhou2023}. Although the use of generative AI and LLMs in particular have shown promise in areas such as content writing, education and training, code generation, healthcare, and many other fields \cite{Ray2023}, the exploration of the capabilities of such LLMs is lacking when it comes to their use in making websites accessible. 


Research on AI in website accessibility is still limited, often focusing on the intersection of universal design learning (UDL) and AI \cite{Kohli2021, Zdravkova2022, Melo2023, SaborioTaylor2024, Hyatt2024}. Recently, Lundqvist et al. \cite{Lundqvist2024} explored how AI can assist UX designers in improving website accessibility for users with diverse needs. They found that while AI enhances design efficiency, it cannot fully automate accessibility compliance; a combination of AI tools and human expertise is necessary to meet standards like WCAG and the European Accessibility Act \cite{DirectiveEU2019}. Othman et al. \cite{Othman2023} investigated ChatGPT's potential in improving web accessibility. They selected two websites, assessed them using WAVE, and used the identified issues for remediation via ChatGPT. They found that ChatGPT can resolve many detected issues, reducing manual effort, but some errors still require manual fixes and human oversight.

Despite these advancements, the highlighted studies are limited to automated evaluations, without incorporating manual assessments, and lack dynamic content. Additionally, previous studies provided only prompts without including screenshots to pinpoint where errors occur. Incorporating screenshots would offer a clearer understanding of the issues and enhance the effectiveness of accessibility evaluations. It also remains unclear whether generative AI tools like ChatGPT inherently prioritize accessibility during code generation and how effectively they address complex, interactive accessibility challenges.

In this study, we use ChatGPT (GPT-4o) \cite{GPT4o2024}, released by OpenAI in May 2024, to generate a fully functional webpage with common website elements and evaluate its accessibility through automated and manual testing. Our goal is to examine GPT-4o's default behavior when generating a webpage without explicitly mentioning accessibility: Does it incorporate accessibility features by default? If so, which features are automatically integrated, and what are the limitations? This investigation is timely, as the 2024 Stack Overflow Developer Survey reports that over 76\% of developers are currently using or planning to use AI coding tools \cite{StackOverflow2024}—a trend unsurprising given the widespread availability of tools like ChatGPT, Claude, and Gemini.

After analyzing GPT-4o's default approach to accessibility, we use GPT-4o itself to fix the identified accessibility errors, assessing its effectiveness in terms of the number of iterations required for both manual and automated issues. We also utilize GPT-4o's visual reasoning capability by providing screenshots of error locations along with prompts. Finally, we present our observations and insights on the challenges and opportunities of using a large language model to design an accessible webpage. To our knowledge, this is the first investigation into ChatGPT's default accessibility behavior and a practical use case of utilizing it to fix existing issues using prompt engineering and its visual reasoning capability.



\section{Methodology}\label{method}



The primary objectives of our study are twofold: first, to evaluate the accessibility of the generated webpage using both manual and tool-based methods, as combining these approaches results in a more comprehensive assessment \cite{alajarmeh2022evaluating}; second, to refine the webpage with the help of ChatGPT (GPT-4o) itself to correct any identified errors, ensuring full compliance with the Web Content Accessibility Guidelines (WCAG) \cite{w3c_wcag_2024}, which serve as the foundational principles for the code.

\subsection{Webpage Design}
We selected a TV series webpage as the focus of our investigation due to its ability to encompass a diverse range of essential website elements commonly found across various web domains. While alternative website types, such as e-commerce platforms, could have been chosen, a TV series webpage provides a representative mix of features, including videos, carousels, pricing tables, images, drop-down menus, and other interactive components. Hence, this selection ensures a comprehensive evaluation of accessibility across a broad spectrum of web design elements, making the findings more generalizable to diverse website types. The prompt we use for the website generation can be seen in Appendix \ref{webpage-prompt}, which contains information about all the webpage sections and their elements.

\begin{figure}
\centering
\includegraphics[width=1\textwidth]{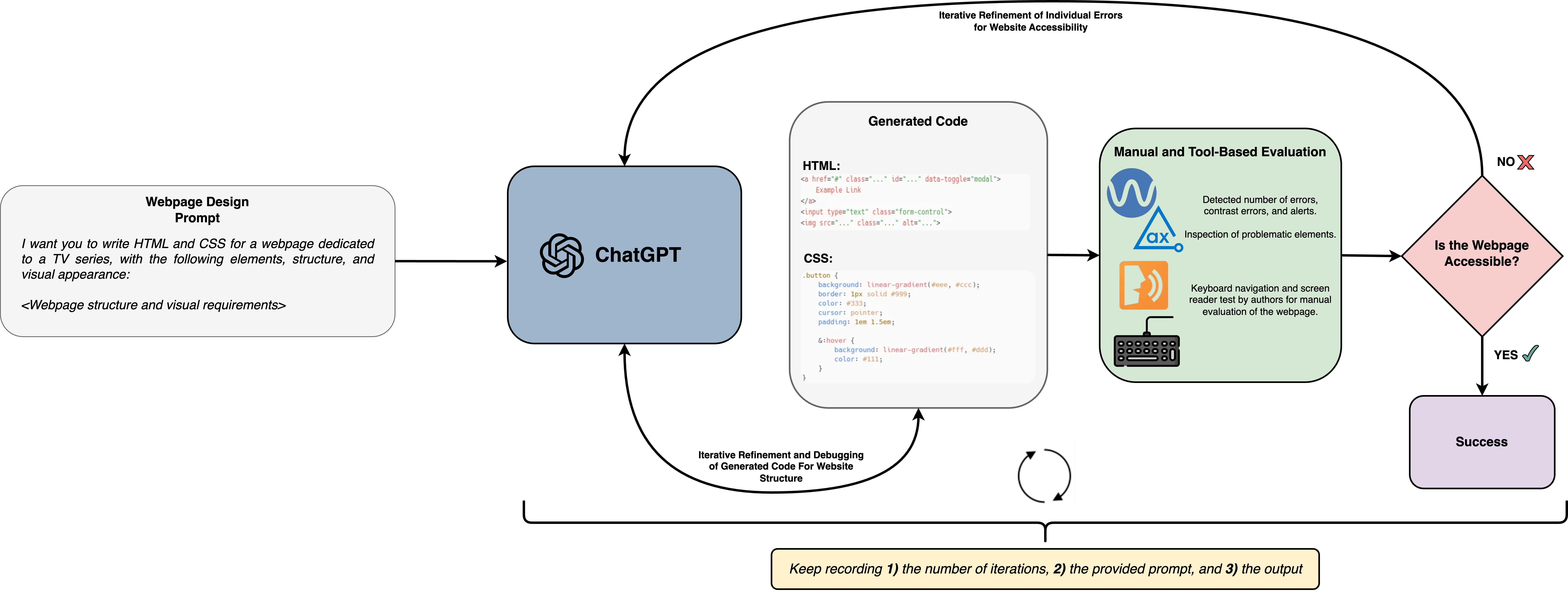}
\caption{Flowchart illustrating the steps involving generating code, evaluating accessibility using manual and tool-based methods, and iteratively refining the code to address identified issues.}
\label{fig:procedure}
\end{figure}

\subsection{Measures}
We employ a mixed-methods approach, using quantitative and qualitative measures to evaluate the generated webpage's accessibility. Quantitative data includes the total number of accessibility issues detected by automated tools WAVE \cite{WAVE2001} and Axe \cite{DequeAxe}, which provide detailed reports. We also record the frequency of each error type (e.g., low contrast, missing form labels, missing navigation). {The qualitative data is gathered through manual testing conducted by the authors, using keyboard navigation and a screen reader extension on Chrome to simulate the experiences of users with disabilities. 

Moreover, we check for various other WCAG guidelines such as whether there are visual indicators? are the complex elements such as dropdowns fully accessible and adhere to WCAG guidelines? does the user have control over time-sensitive events? are the tables accessible? and so on. Therefore, manual analysis is more thorough than automated testing, as it goes beyond analyzing the source code. Importantly, we document the number of iterations (feedback prompts) required to fix each detected accessibility error, serving as a metric to assess ChatGPT's efficiency and adaptability in iteratively refining the webpage to achieve WCAG 2.2 AA compliance. 



We also use a weighted average that accounts for the complexity of each task. Since some errors took more iterations to be fixed than others. The weighted average of iterations provides a more nuanced measure than a simple average by giving more weight to complex tasks. We define the weighted average based on the following formula:
\[
\text{Weighted Average Iterations (WAI)} = \frac{\sum (x_i \times w_i)}{\sum w_i}
\]
where \( x_i \) is the number of iterations required to solve a particular issue \( i \), \( w_i \) is the weight assigned based on the complexity of the issue, and \( \sum w_i \) is the total sum of the weights. In this approach, we categorize issues based on their iteration count:
\begin{itemize}
    \item \textbf{Complex Tasks}: Issues requiring more than a certain threshold of iterations (e.g., more than 10 iterations) are considered complex and assigned a higher weight (e.g., \( w = 2 \)).
    \item \textbf{Simple Tasks}: Issues requiring fewer iterations (e.g., less than or equal to 10 iterations) are considered simpler and assigned a lower weight (e.g., \( w = 1 \)).
\end{itemize}

Finally, we compute the accuracy by looking at the total number of errors ChatGPT was able to resolve (NER) compared to the total detected errors (TED), as shown below:
\begin{equation}
\text{Accuracy} = \frac{\text{NER}}{\text{TED}} \times 100
\end{equation}

\subsection{Procedure}

To begin, we provided ChatGPT with detailed instructions on the desired webpage structure and visual layout (see Appendix \ref{webpage-prompt}). We used the pro version of ChatGPT (GPT-4o) on a newly created account to ensure no prior conversation history influenced its responses, and the newly introduced "memory" feature was turned off to prevent knowledge transfer between chat sessions. ChatGPT was chosen for this study due to its superior performance in recent human-evaluated code generation benchmarks compared to other prominent LLMs like Claude or Gemini \cite{GPT4o2024}, and because it has become widely used by developers worldwide \cite{Team2023}. While ChatGPT maintains context within a chat session, the context window of LLMs is limited, and as conversations grow longer, they start to forget earlier interactions \cite{an2024make}. To mitigate this limitation, we either provided the relevant code again to refresh its context or started a new chat session and provided the context anew.



We then checked the webpage’s structure and visual alignment against the given requirements described in the prompt. By ``structure'' we mean are all the required elements such as the navigation bar, buttons, forms, headers, etc. present in the webpage? Do they follow the structural hierarchy that we specified in the prompt? By ``visual alignment'' we mean do the elements have the required color scheme? are the font styles consistent? do buttons have the required appearance? And so on. Lastly, we will check whether the generated code is free of any errors. If the webpage did not fully meet our expectations, we prompted it again providing the issues that were present, asking it to fix the issues, and specifying what we needed instead. This process continued until the whole webpage design and structure were satisfactory.

After completing the webpage design, we conducted tool-based and manual accessibility evaluations, in that order. For tool-based testing, we used WAVE and Axe to identify accessibility issues, examining the frequency and total number of errors and alerts. Each error was analyzed to pinpoint its location on the webpage. Depending on the error, we either: a) prompted ChatGPT to fix it, b) provided the HTML/CSS code snippet for correction, or c) supplied a screenshot of the error location for ChatGPT to address. The output from the ChatGPT containing the updated code is then used, without any manual intervention or review. If the error persisted, we provided the feedback again. This process continued until all detected errors were resolved. Issues found during the manual evaluation were addressed similarly. If ChatGPT was unable to fix an issue after multiple attempts with diminishing returns, we noted it and concluded that ChatGPT could not resolve it.

We began fixing all accessibility errors in a new chat session, providing ChatGPT with the initial code and WCAG 2.2 guidelines \cite{w3c_wcag_2024} to ensure familiarity with accessibility standards from the outset, as sufficient background and context in the prompt increases the likelihood of accurate code generation \cite{white2024chatgpt, liu2024no}. This entire process is illustrated in Figure \ref{fig:procedure}. Iteratively refining the web pages based on feedback mirrors real-world development practices and leverages ChatGPT’s ability to adjust outputs based on new prompts. In this study, we do adhere to best prompt engineering practices for code generation as outlined by White et al. \cite{white2024chatgpt}, although, we also focus on how a typical developer might interact with ChatGPT, rather than as a prompt engineering expert.


\section{Results}\label{results}
\begin{figure}
\centering
\includegraphics[width=1\textwidth]{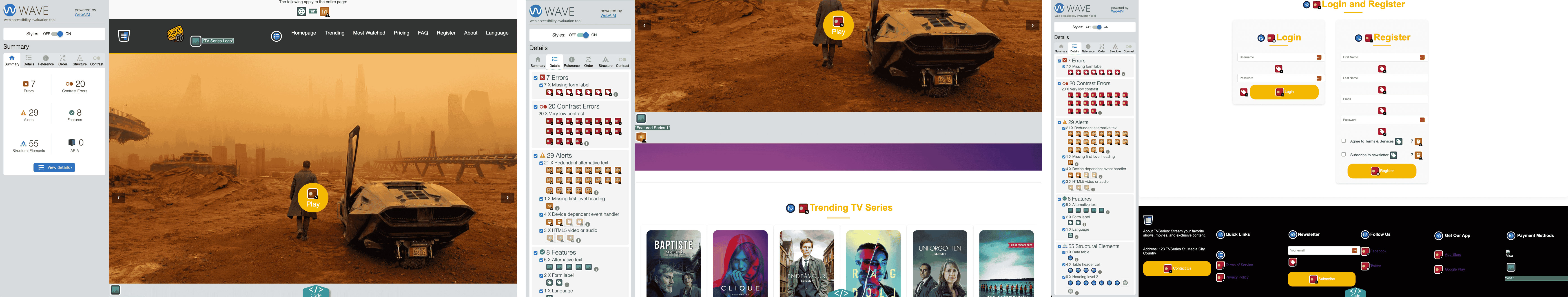}
\caption{Automated errors detected by WAVE in the initial design. This excludes text contrast errors from the drop-down options and modals.}
\label{fig:auto1}
\end{figure}


\subsection{Positive Accessibility Features in the Default Generated Code}
After the initial prompt, ChatGPT demonstrated some adherence to accessibility best practices in the generated webpage code. 
The generated code consistently used semantic tags over generic \textit{div} elements. This included appropriate tags such as \textit{nav} for navigation, \textit{section} for various page sections, and \textit{footer} for the footer, and so on. The \textit{lang} attribute was set to “en” (English), helping screen readers interpret the page language, along with a descriptive \textit{title} tag, which is the first thing the screen readers say. Headings were organized hierarchically, with \textit{h3} nested within \textit{h2} tags. Links were visually distinguished with underlines and color, differentiating them from non-link text. The generated code clearly distinguished between links (which navigate) and buttons (which trigger actions). The \textit{alt} attribute was provided for all \textit{img} tags, either with a placeholder or left empty when appropriate, leaving room for content-specific alt text. Elements with semantic tags followed a simple and consistent navigation order. The forms had one input field per line, which is considered a best practice, instead of multiple input fields in a single line. In terms of visual accessibility, the text-heavy sections with important content, such as all the headings, frequently asked questions, and login and registration forms, were aligned to the left, following the F-pattern \cite{Pernice2017} commonly observed in left-to-right reading cultures. Finally, ChatGPT also defaulted to a simple, accessible font style “Arial, sans-serif”. The webpage also loosely followed a Z-pattern \cite{Babich2017}, a typical layout pattern where critical elements are arranged along a “Z” shape, guiding the eye from the logo on the top left to the navigation on the top right, down to the content on the left side, and ending at the call-to-action on the bottom right (in this case, the newsletter subscription).

\begin{table}[H]
\caption{Contrast Ratios and WCAG Compliance}
\centering
\begin{tabular}{|l|c|c|c|}
\hline
\textbf{Element} & \textbf{Text Contrast Ratio} & \textbf{Frequency} & \textbf{WCAG 2.2 AA Level} \\ \hline
Form headings & 1.71 & 2 & Not Compliant \\ \hline
Footer links & 2.23 & 6 & Not Compliant \\ \hline
Buttons & 1.8 & 5 & Not Compliant \\ \hline
Section headings & 1.8 & 5 & Not Compliant \\ \hline
Modals & 1 & 2 & Not Compliant \\ \hline
Dropdown options hover & 1.8 & 6 & Not Compliant \\ \hline
Navigation menu items & 12.63 & 8 & Compliant \\ \hline
Movie posters & 12.63 & 21 & Compliant \\ \hline
Table rows & 12.63 & 11 & Compliant \\ \hline
FAQ items & 12.63 & 10 & Compliant \\ \hline
Footer & 21 & 6 & Compliant \\ \hline
Checkboxes in form & 12 & 2 & Compliant \\ \hline
\end{tabular}

\label{tab:contrast}
\end{table}

\subsection{Accessibility Limitations in the Default Generated Code}
 Although ChatGPT used semantic tags by default, it failed to include a “main” landmark. Consequently, it did not provide a “skip-to-main content” link, a feature for allowing screen reader users to quickly skip to the main content of the webpage. The absence of a level-one heading (\textit{h1}) was another oversight. 
 One of the most common accessibility issues we identified was low contrast across various elements. This indicates that the contrast issue is a recurrent problem in the default webpage generated by ChatGPT. While links were visually distinct (underlined and colored) as mentioned before, the contrast was far below WCAG 2.2 level AA standards. Low contrast was also noted in headings, buttons, and hover text. The contrast ratios of various elements and their compliance with WCAG 2.2 AA level are shown in Table \ref{tab:contrast} along with an illustration in Figure \ref{fig:dist_wacg}. Accessibility best practices dictate that a website should be fully navigable via keyboard, yet ChatGPT’s generated page lacked this functionality. Many sections, including the “FAQ” section, tables, modals, and interactive content like “trending” carousel and “most watched TV shows” were inaccessible to keyboard navigation. This was because there was no \textit{tabindex} applied to non-semantic elements, leaving some items out of the natural navigation flow. 
 
 The login and registration forms did not include labels for input fields. Placeholder text was present, but placeholders should ideally provide examples rather than act as field labels. Link descriptions were overly brief and could be confusing. For instance, rather than saying “Privacy Policy” a more accessible option would be “About Our Privacy Policy”. 
 The table on the page lacked a caption or summary, crucial for context in tabular data. Additionally, the table headers did not use the \textit{scope} attribute to define the column headers, which hinders screen reader interpretation. Drop-down options were entirely inaccessible via keyboard, and they suffered from low contrast. Modals for the registration form were not opening at first using the keyboard, when this was fixed, the focus was not shifted to the modal content, leaving it on the page beneath the modal. The hero section included a video with a \textit{video} tag but ChatGPT did not suggest or give a code snippet to add captions. Finally, Although the initial prompt requested a moving carousel, ChatGPT did not include a “skip” option to bypass the carousel content, nor did it alert screen reader users when the carousel content changed. This oversight limits accessibility for those who rely on screen readers and creates an obstacle in navigating lengthy content. The top 5 "common" errors and their frequency on the generated webpage are shown in Figure \ref{fig:common}. The automated errors detected by WAVE are shown in Figure \ref{fig:auto1}, whereas the Axe report is shown in Figure \ref{fig:axe}.

\begin{table}[H]
\caption{Automated Detected Errors (WAVE + axe) and Their Resolution}
\centering
\begin{tabular}{|l|c|c|}
\hline
\textbf{Error Description} & \textbf{Fully Resolved} & \textbf{Iterations} \\ \hline
Missing form labels (forms) & Yes & 3 \\ \hline
Missing form label (footer) & Yes & 1 \\ \hline
No main landmark & Yes & 1 \\ \hline
No level one heading & Yes & 2 \\ \hline
Low contrast text (play button) & Yes & 1 \\ \hline
Low contrast text (forms) & Yes & 1 \\ \hline
Low contrast text (form modals) & Yes & 2 \\ \hline
Low contrast text (section headings) & Yes & 2 \\ \hline
Low contrast text (buttons in footer) & Yes & 1 \\ \hline
Low contrast text (buttons in forms) & Yes & 1 \\ \hline
Low contrast text (headings in footer) & Yes & 1 \\ \hline
Low contrast text (links in footer) & Yes & 1 \\ \hline
Device dependent event handler & Yes & 1 \\ \hline
\multicolumn{2}{|l|}{\textbf{Total Iterations}} & \textbf{18} \\ \hline
\multicolumn{2}{|l|}{\textbf{WAI}} & \textbf{1.4} \\ \hline
\multicolumn{2}{|l|}{\textbf{Accuracy}} & \textbf{100\%} \\ \hline
\end{tabular}
\label{table:auto}
\end{table}
\begin{table}[H]
\caption{Summary of Qualitative Errors and Their Resolution}
\centering
\begin{tabular}{|l|c|c|}
\hline
\textbf{Error Description} & \textbf{Fully Resolved} & \textbf{Iterations} \\ \hline
No table accessibility & No & 36 \\ \hline
No hero image visual indicator & No & 32 \\ \hline
No FAQ items navigation & Yes & 32 \\ \hline
No drop-down accessibility & Yes & 28 \\ \hline
Proper alt text missing & Yes & 22 \\ \hline
No navigation bar items navigation & Yes & 16 \\ \hline
Carousel play/pause button missing & Yes & 11 \\ \hline
Play button (Missing ARIA label + icon + visual indicator) & Yes & 10 \\ \hline
No carousel items navigation & Yes & 9 \\ \hline
No form modals accessibility (navigation + opening) & Yes & 7 \\ \hline
No skip-to-main-content link & Yes & 2 \\ \hline
Most watched items navigation missing & Yes & 2 \\ \hline
All headings inaccessible & Yes & 2 \\ \hline
Footer payment methods (missing links, aria labels) & Yes & 3 \\ \hline
Logo navigation missing & Yes & 1 \\ \hline
Carousel auto-disable button missing & Yes & 1 \\ \hline
Skip-carousel-link missing & Yes & 1 \\ \hline
Alert carousel content change missing & Yes & 1 \\ \hline
Hero image inaccessible & Yes & 1 \\ \hline
Footer links, text content navigation missing & Yes & 1 \\ \hline
Missing underlines on follow and quick links in the footer & Yes & 1 \\ \hline
Videos have no captions & Yes & 1 \\ \hline
\multicolumn{2}{|l|}{\textbf{Total Iterations}} & \textbf{220} \\ \hline
\multicolumn{2}{|l|}{\textbf{WAI}} & \textbf{13.4} \\ \hline
\multicolumn{2}{|l|}{\textbf{Accuracy}} & \textbf{90.91\%} \\ \hline
\end{tabular}
\label{table:manual}
\end{table}
\subsection{Effectiveness of ChatGPT in Understanding Accessibility Requirements and Making The Webpage Accessible}
 In this section, we examine ChatGPT's ability to fix detected accessibility issues categorized into automated and manually identified issues, shown in Tables \ref{table:auto} and \ref{table:manual}, respectively, along with the number of iterations (feedback prompts), and their resolution. When it came to automated errors, most issues were resolved in one iteration. However, some issues like “Missing form labels (forms)” and “No level-one heading” took more than one iteration to resolve. The reason it took 3 iterations to fix the missing form labels issue is that ChatGPT provided the labels on a separate line instead of beside the associated input field, which is often the best practice. Contrast errors typically required only one iteration to fix. The table’s total number of iterations to fix all automated errors is relatively low (18 iterations in total).

 \begin{figure}[H]
\centering
\includegraphics[width=0.7\textwidth]{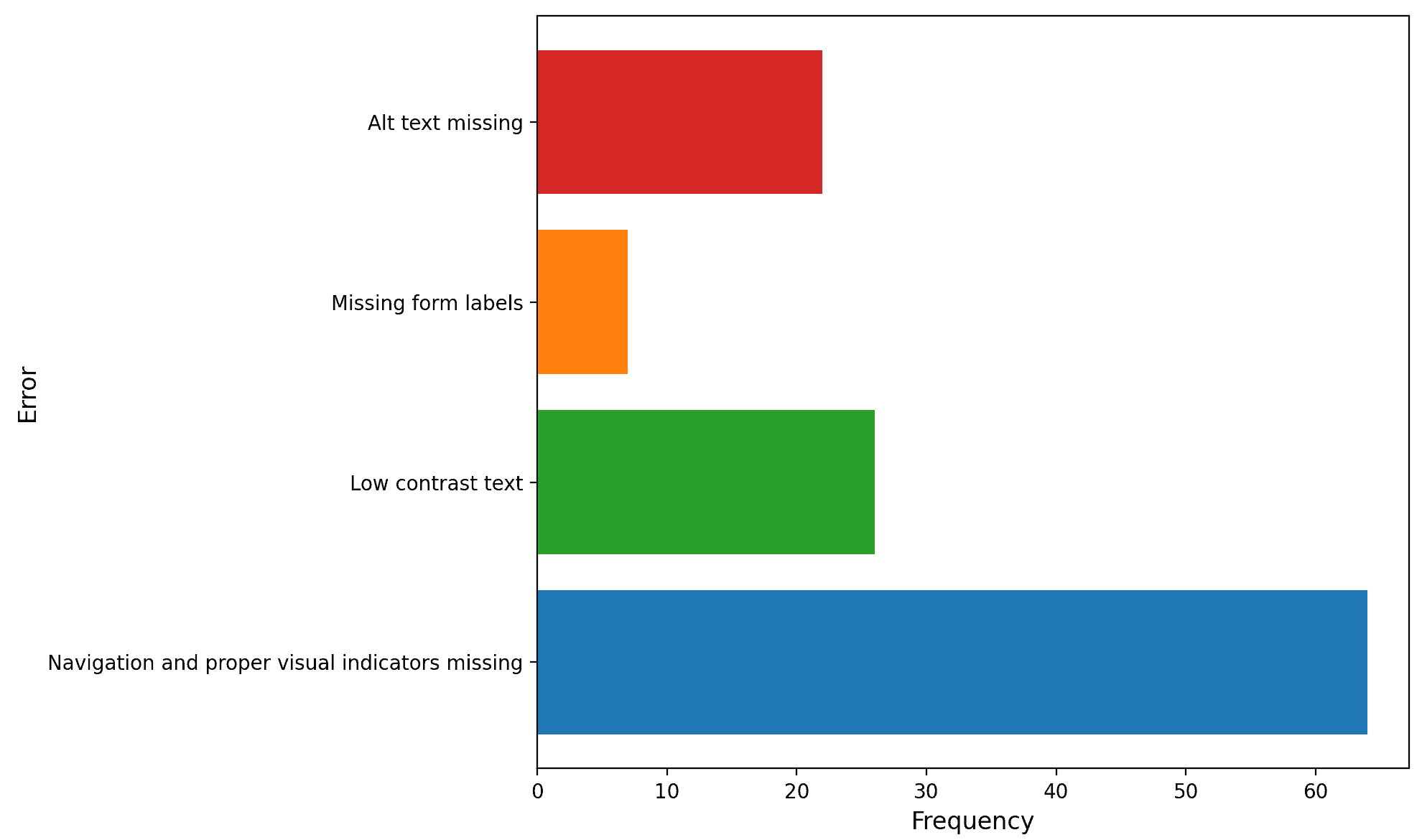}
\caption{Frequency of common errors in the ChatGPT-generated webpage.}
\label{fig:common}
\end{figure}

Manual errors, on the other hand, highlighted areas where ChatGPT struggles due to the need for a deeper understanding of context and user-centered design principles. Some errors were straightforward to fix, such as creating a “skip-to-main-content” link, fixing keyboard navigation on static items, and adding a “skip-carousel” link. However, more complex issues required significantly more iterations. For example, making the drop-down menu fully accessible per WCAG 2.2 guidelines took 28 iterations (see Figure \ref{fig:dropdown} for the checklist). Even after making it keyboard-accessible, the screen reader focused on the entire drop-down window instead of the specific item, failing to announce selections, which required additional iterations. The FAQ section was another challenge. Initially, items expanded on mouse hover but needed keyboard navigation. Achieving both functionalities took 32 iterations, as ChatGPT struggled to implement them simultaneously. Adding a functional play/pause button to the carousel was also complex. Initially nonfunctional due to JavaScript issues, further adjustments made it navigable and properly aligned. Making the modals in the registration form accessible required 7 iterations, involving keyboard navigation, opening via keyboard, shifting focus to the close button upon opening, and preventing focus on elements behind the modal.


Making the table fully accessible required the most iterations, as we had to ensure each row and cell was keyboard-navigable and screen reader accessible. Despite making the table nearly fully accessible per WCAG 2.2 guidelines (see Figure \ref{fig:table} for the checklist), in addition to making rows navigable using the tab key, one bug persisted: when we pressed tab while being on the last row, it refocused on the first cell of that row instead of moving to the next webpage element. Despite our efforts and ChatGPT's JavaScript suggestions, we could not fix this final issue. Similarly, the "No hero image visual indicator" error posed challenges. Despite multiple attempts, ChatGPT couldn't properly apply a focus visual indicator around the entire border of the image using CSS. The indicator appeared only on the top and bottom or sometimes right and left, but it couldn't fix it entirely. 

Making the carousel accessible was challenging; it needed to pause when a user focused on an item, support keyboard navigation, and resume when focus was removed. Despite ChatGPT initially handling navigation bar items by default, the error persisted because accessibility involves more than tab navigation—we had to enable arrow key navigation, allow activation with enter or space, and support typing letters to focus on matching items. Additionally, making the drop-down accessible disrupted other navigation items, requiring further iterations to ensure the drop-down could be navigated without affecting the rest of the navigation functionality.


Encountering the error “No proper alt text for images,” we wondered if ChatGPT-4o's visual reasoning could generate descriptive alternative text. It could. After providing guidelines—to be concise (1-2 sentences), focus on key elements, end with a period, and avoid starting with phrases like “image of” or “picture of”—we submitted all 22 webpage images to ChatGPT. The resulting text was impressively satisfactory after just one iteration, even without an in-depth evaluation of its descriptive quality. An example is shown in Figure \ref{fig:alt-example}.

The webpage’s visual appearance, before and after implementing accessibility improvements, is shown in Figures \ref{fig:before1} and \ref{fig:before2}. While many of the accessibility issues were inherent rather than purely visual, noticeable changes have been made. These include enhanced contrast, the addition of icons to the play button, improved button design, the inclusion of form labels, a fully refined footer, better spacing between FAQ items, a polished table layout, the addition of a level-one heading, and a more accessible play button.


\section{Discussion}\label{discussion}
The present study reveals that while ChatGPT does not inherently produce fully accessible webpage code, it can address accessibility issues when prompted, achieving an accuracy of 90.91\% in resolving manual errors. Nevertheless, human expertise and review remains crucial in the development process and developers cannot rely solely on ChatGPT to produce fully accessible code By default, as its generated code often includes WCAG violations, often relating to contrast, visual aids, and navigation, likely reflecting the inaccessibility of publicly available training data and a general lack of awareness about accessibility in web development \cite{WebAIM2024, Freire2008}.

We found that ChatGPT proved effective at fixing straightforward, automated accessibility errors but struggled with complex, manually detected issues, often requiring additional guidance and clarification, consistent with findings by \citeauthor{Othman2023} \cite{Othman2023}. The iterations increased as the issues became more complex, this trend is illustrated in Figure \ref{fig:rising}. This ties into another limitation which is that when it came to resolving errors that required many rules to be implemented such as a drop-down, ChatGPT required more iterations. This suggests that complex components involving extensive rules or JavaScript functionality demand more prompts and iterations from ChatGPT than simpler, static components.

\begin{figure}
\centering
\includegraphics[width=0.3\textwidth]{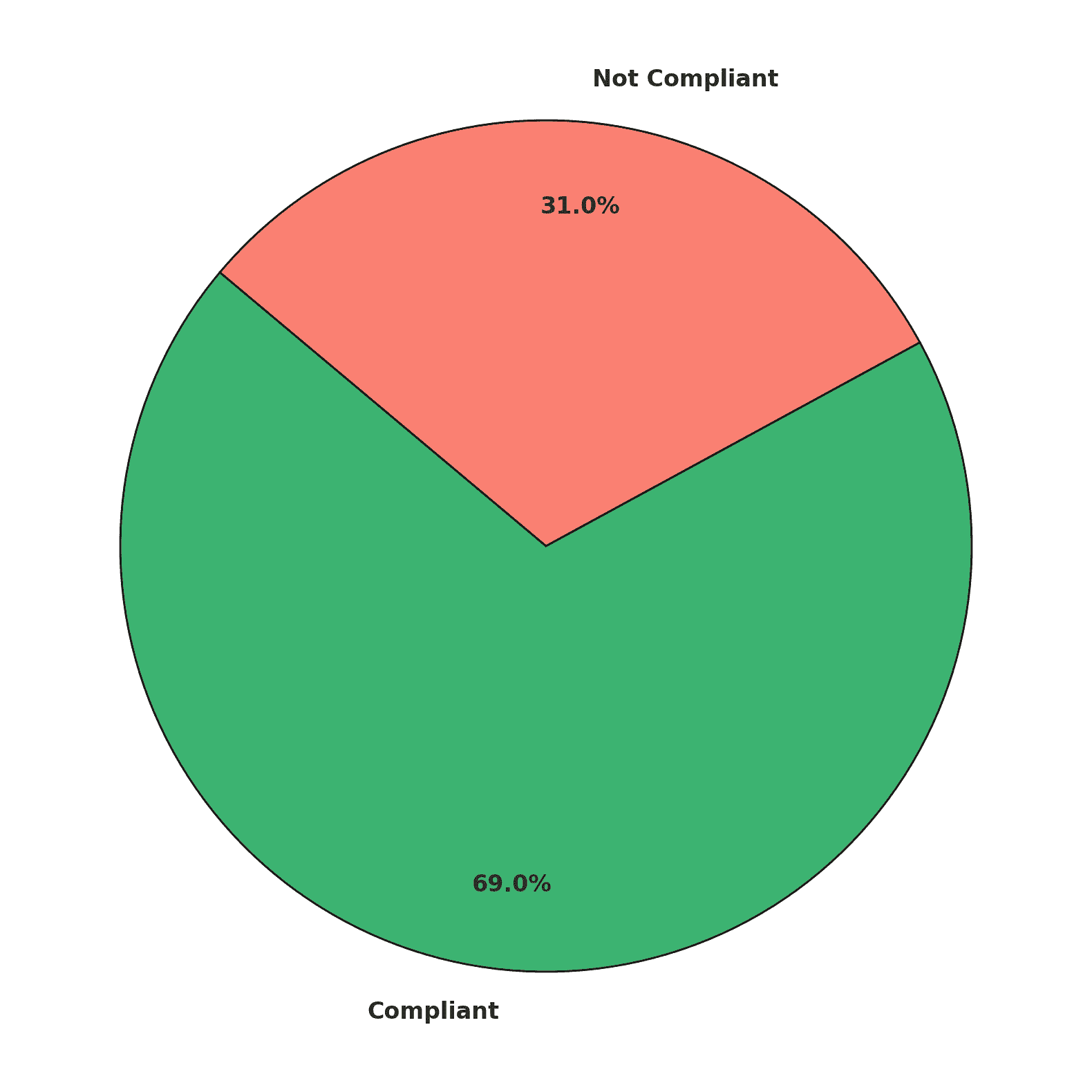}
\caption{Distribution of webpage elements by compliance with WCAG 2.2 text contrast ratios.}
\label{fig:dist_wacg}
\end{figure}

As mentioned before, fixing one component could potentially lead to breaking another component, especially when the former component requires fixes in the JavaScript code. Therefore, ChatGPT may inadvertently compromise functionality in others, necessitating further adjustments to restore overall usability and accessibility. This is partly because ChatGPT often included unnecessary code for unrelated components, disrupting adjacent functionality. Adding instructions like “Provide only the relevant code” resolved this. Explicitly telling ChatGPT not to alter certain functionalities or providing the working code ensured new features didn't compromise existing ones. Debugging by logging console values, sharing outputs with ChatGPT, and receiving effective fixes was also helpful. Prompting it to “Take a different approach” reduced iterations and avoided repetitive outputs, though it didn't always resolve issues immediately.

 Even after providing the full WCAG 2.2 guidelines (over 13,800 words), ChatGPT did not consistently apply them and required specific instructions for each task. This suggests that as conversations expand, ChatGPT struggles to retain lengthy prompts, in line with An et al. \cite{an2024make}, and benefits more from concise, targeted instructions. This has practical implications for how developers should interact with AI tools: breaking down tasks into smaller, manageable pieces can enhance efficiency and reduce frustration. Reintroducing the full webpage code during extended discussions also improved performance when responses began to lag or lose coherence, although the exact threshold for this reset is unclear. 

A significant observation was the importance of clear, structured feedback, in line with White et al. \cite{white2023prompt}. Including three key points in the prompt—(1) the action taken, (2) the expected outcome, and (3) the actual result—was more effective than generic retry requests. Combining screenshots with detailed textual prompts yielded better results, as visual context enhanced ChatGPT’s ability to address issues like contrast adjustments. For example, a screenshot helped it account for a component's background color, leading to a more accurate solution. Lastly, we also experimented with having ChatGPT refine our detailed prompts for clarity. In another session, we submitted a complex prompt and asked ChatGPT to simplify it for improved readability. We found that using this “refined” prompt led to better results. While we didn’t adopt this approach consistently, it proved effective when we did.

\begin{figure}[H]
\centering
\includegraphics[width=1\textwidth]{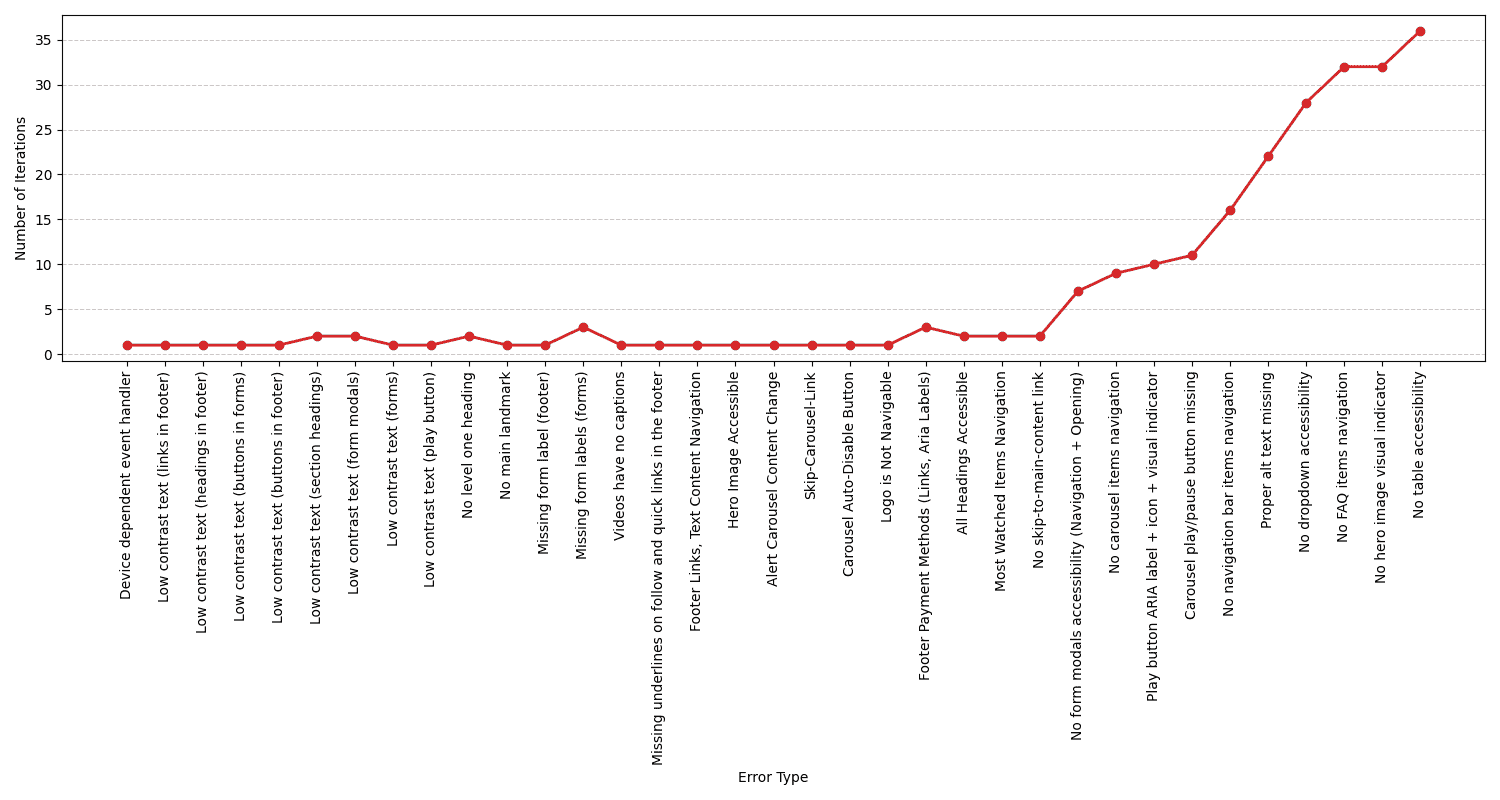}
\caption{Iterations vs. error types. The number of iterations increase as the complexity of error types increase.}
\label{fig:rising}
\end{figure}

\section{Limitations and Future Work} \label{limitations}
Our study focused exclusively on the ChatGPT-4o model, not other proprietary or open-source LLMs. Expanding the analysis to include other models could provide insights into their accessibility strengths and limitations. Moreover, our results heavily depend on the quality of the prompts used. Findings may differ with less refined or more concise prompts, as prompt engineering introduces variability. Investigating systematic methods for optimizing prompts could lead to more consistent outputs from LLMs. This study primarily focused on how typical developers might interact with ChatGPT, rather than as prompt engineering experts. Lastly, manual testing and evaluation of qualitative features like user experience were based on the authors' judgment, introducing subjectivity that might influence the results.

\section{Conclusion}\label{conclusion}


This study demonstrates that while ChatGPT can address a range of accessibility issues in web development—from straightforward errors to more complex challenges—it does not inherently prioritize accessibility in its default code generation. Developers can leverage ChatGPT as a supportive tool to assist with coding tasks and improve accessibility features, but it should not replace human judgment and expertise. The insights and best practices identified offer practical strategies for developers aiming to create accessible websites more effectively. Our findings also provide researchers with a clearer understanding of the current limitations of large language models like ChatGPT, highlighting areas for enhancement through improved training data focused on accessibility compliance. Continued efforts to integrate AI tools with human expertise are essential. By collaborating closely with accessibility professionals, developers can harness the potential of AI while ensuring that web content remains inclusive and compliant with accessibility standards.

\bibliographystyle{ACM-Reference-Format}
\bibliography{sample-base}

\appendix 
\section*{Appendix}

\section*{Detailed Prompt for TV Series Webpage Design} \label{webpage-prompt}

\begin{quote}
\small
\raggedright
I want you to write HTML and CSS for a webpage dedicated to a TV series, with the following elements, structure, and visual appearance:

The webpage should include multimedia elements (images, videos, audio), interactive components (forms, links, tables, dropdowns, modals), and dynamic content. The navigation bar and footer should have a dark background, while the rest of the website should have a white background. Buttons should have a dark yellow color scheme (\#f5b700).

1. \textbf{Navigation Bar}: A logo should be aligned to the left, and menu items (“Homepage,” “Trending,” “Most Watched,” “Pricing,” “FAQ,” “Register,” “About,” and “Language” (dropdown with options for English, Norwegian, and Hindi)) should align to the right.

2. \textbf{Hero Section}: A full-width thumbnail image should represent a video with an overlaid “Play” button. When the “Play” button is clicked, the image, button, and text should disappear, and a video (taking the entire hero section width) should play. Add a manual slider in the Hero Section that rotates featured series, triggered only by left or right arrow clicks, keeping the “Play” button fixed as content changes.

3. \textbf{Trending TV Series Section}: A grid layout showcasing trending TV series. Display a total of 12 items, showing the first 6 initially and automatically swiping to the right to reveal the other 6 after a 4-second delay. This left-right swiping should loop continuously.

4. \textbf{Most Watched Section}: A grid layout with 3 rows, each row showing 3 movie posters.

5. \textbf{Subscription Plans}: A table showing subscription options (Free, Plus, and Pro tiers), highlighting features like the number of devices, video quality, and exclusive content.

6. \textbf{FAQ Section}: A Frequently Asked Questions section with collapsible headings for each question.

7. \textbf{Login and Registration Forms}: 
   - \textbf{Login}: Fields for “Username” and “Password.”
   - \textbf{Registration}: Fields for “First Name,” “Last Name,” “Email,” “Password,” and two checkboxes for newsletters and terms. A clickable “i” icon by the terms checkbox should trigger a pop-up displaying terms and conditions.

8. \textbf{Footer}: The footer should include several sections:
   - \textbf{About Us (Left)}: Brief service description and company address, with a Contact button.
   - \textbf{Quick Links (Middle-Left)}: Links to Terms of Service, Privacy Policy, Help Center, Contact Us, FAQ, Careers, Blog, and Affiliate Program.
   - \textbf{Newsletter Subscription (Bottom-Center)}: An email input box and Subscribe button.
   - \textbf{Social Media Links (Bottom-Left)}: Links to Facebook, Twitter, Instagram, YouTube, LinkedIn, and Pinterest, with an email contact below.
   - \textbf{Get Our App (Middle-Right)}: Links to download the app from Apple’s App Store and Google Play Store.
   - \textbf{Payment Methods (Far Right)}: Icons for Visa, MasterCard, PayPal, American Express, and Bitcoin.
\end{quote}
\normalsize

\begin{figure}[h!]
\centering
\includegraphics[width=0.7\textwidth]{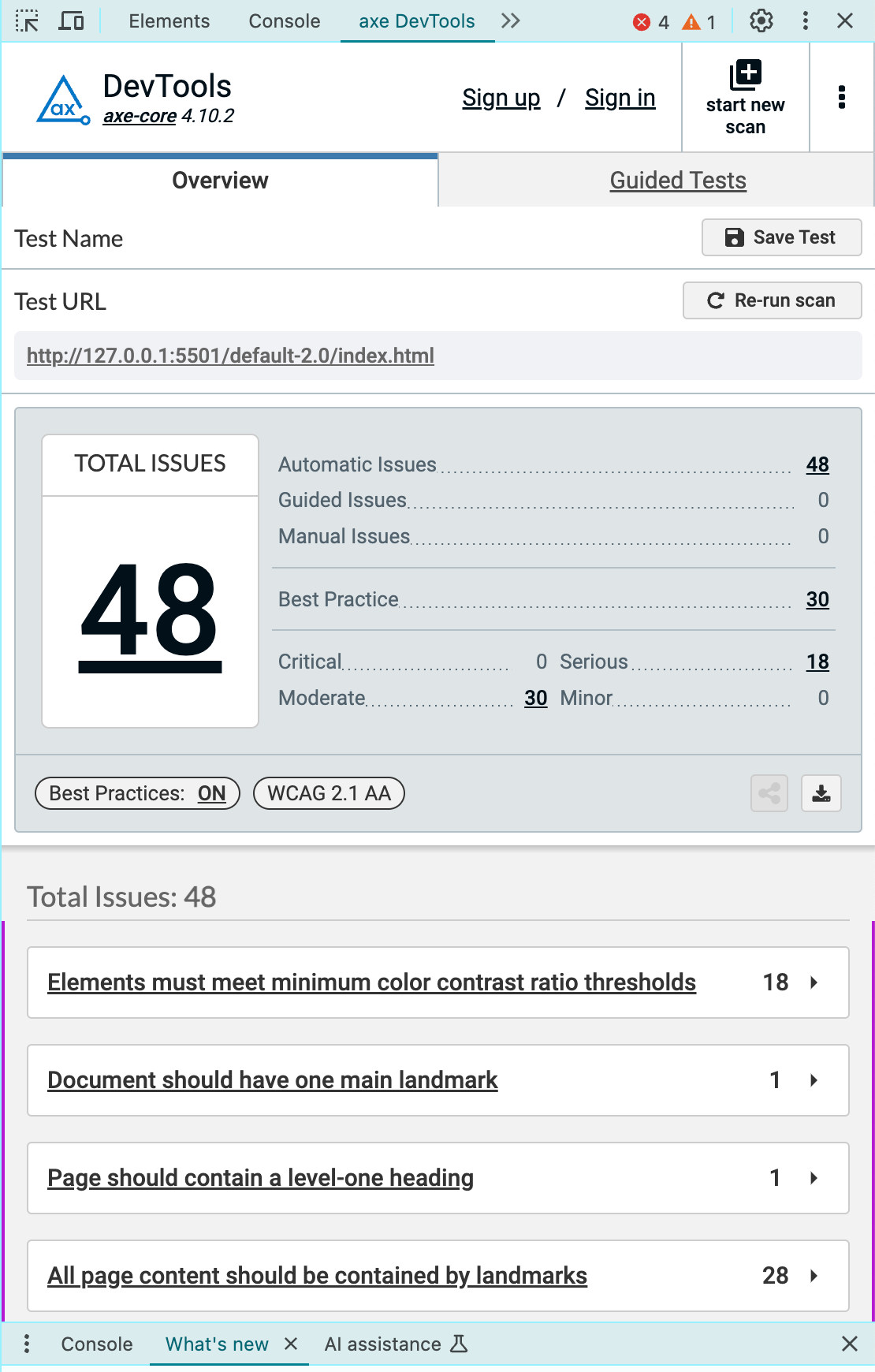}
\caption{Automated errors detected by Axe in the initial design.}
\label{fig:axe}
\end{figure}

\begin{figure}[h!]
\centering
\includegraphics[width=1\textwidth]{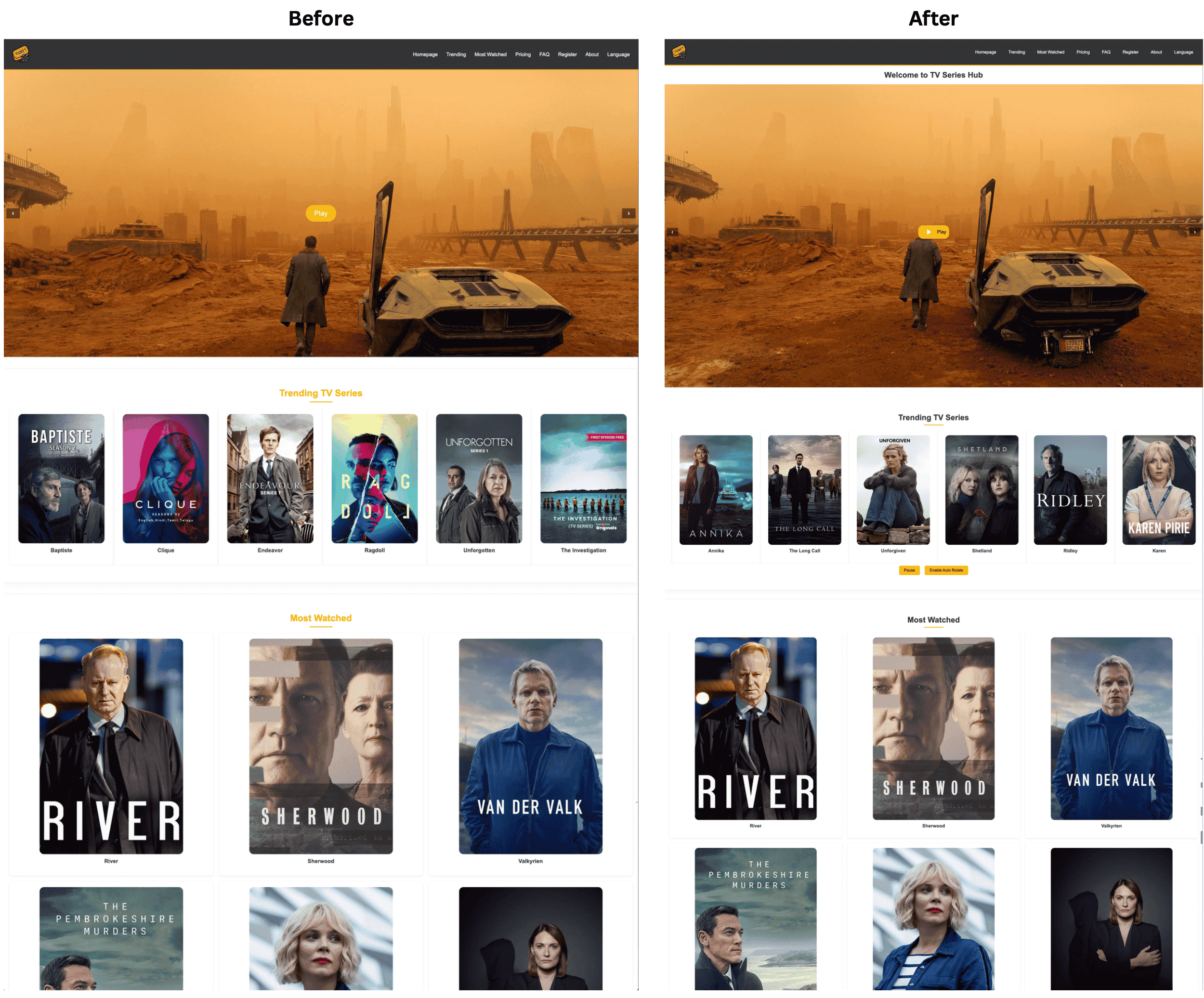}
\caption{Before vs. after adding accessibility elements in the webpage.}
\label{fig:before1}
\end{figure}

\begin{figure}[h!]
\centering
\includegraphics[width=1\textwidth]{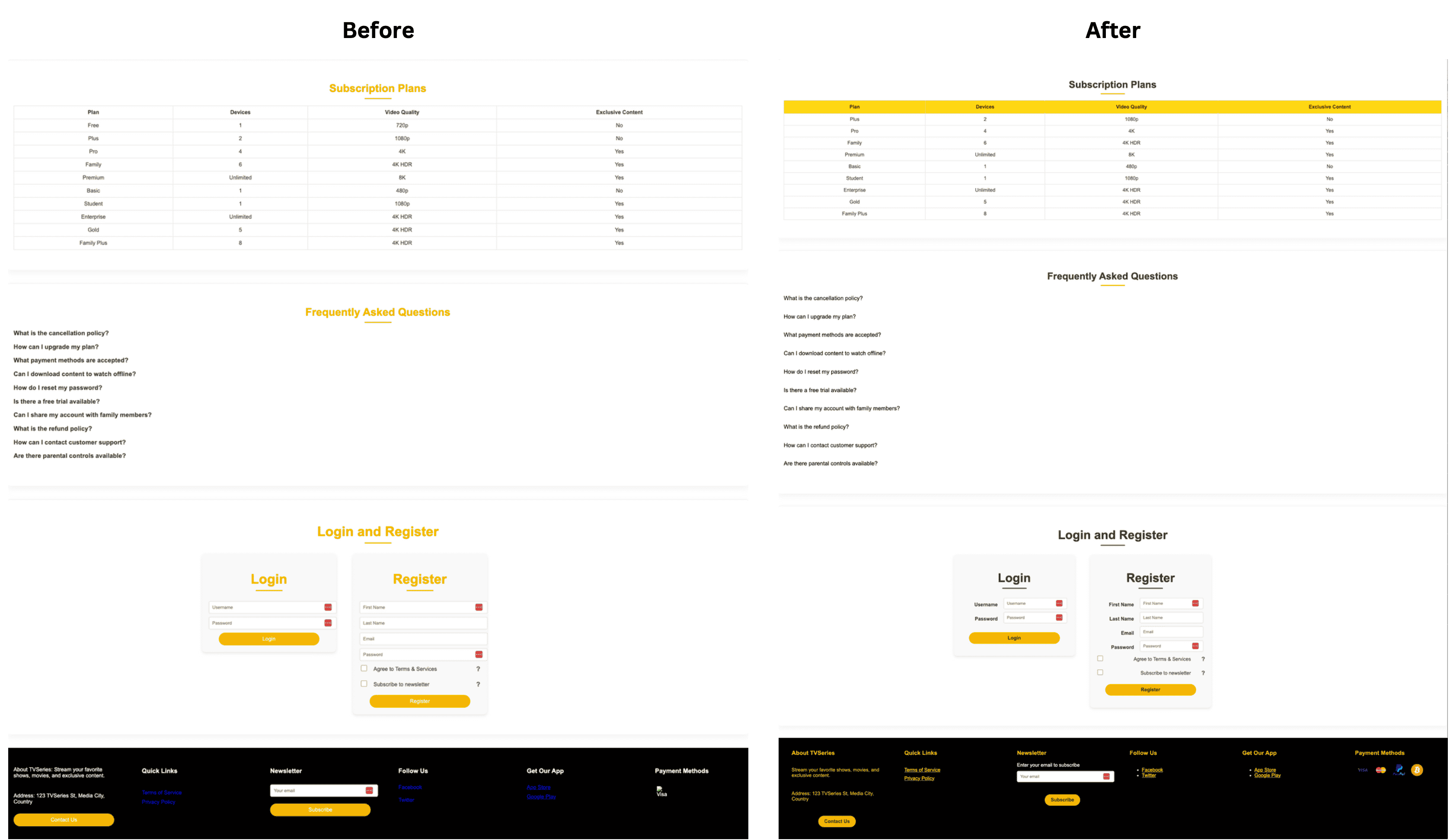}
\caption{Before vs. after adding accessibility elements in the rest of the webpage.}
\label{fig:before2}
\end{figure}

\begin{figure}[h!]
\centering
\includegraphics[width=0.5\linewidth]{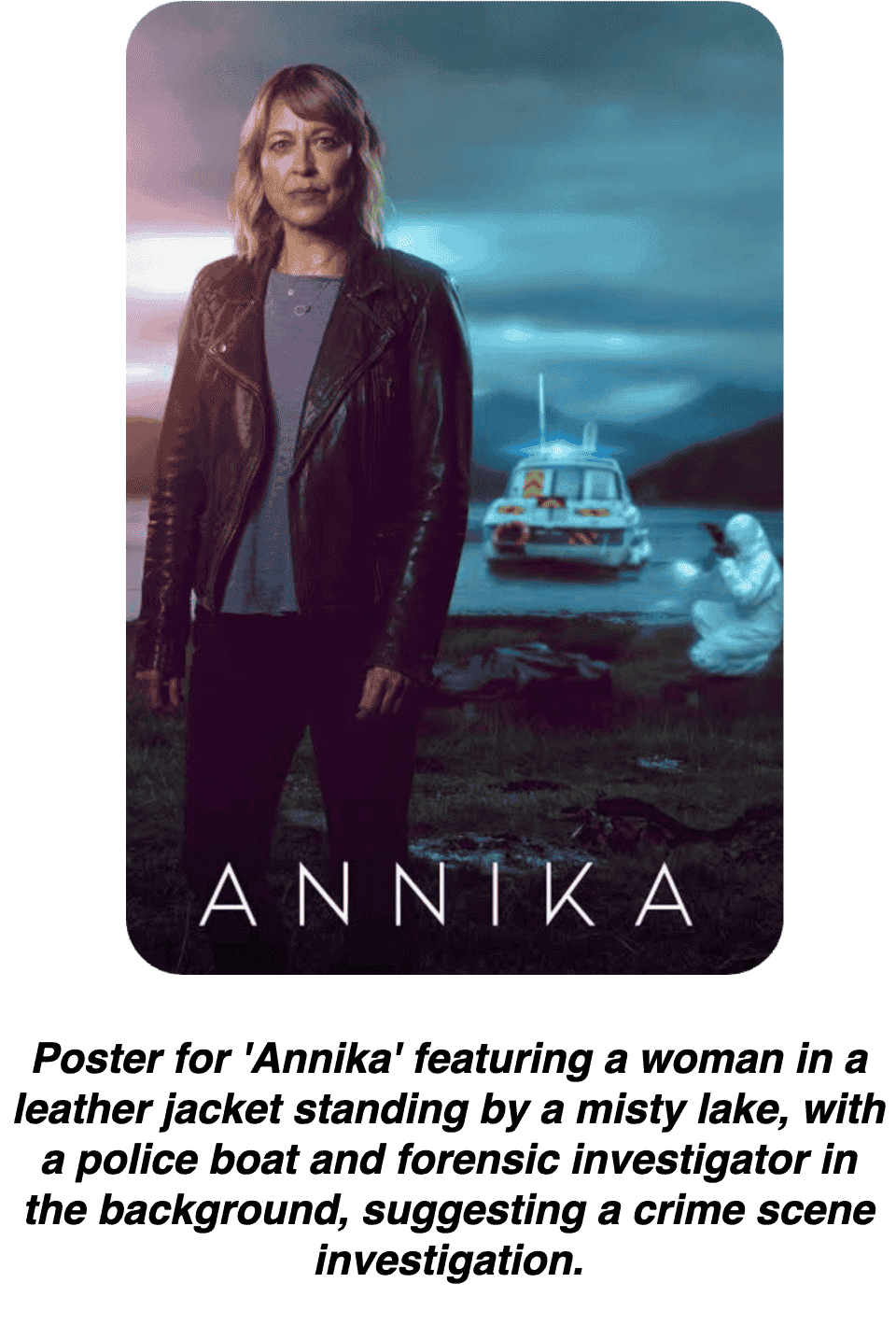}
\caption{An example of an alternative text generated by ChatGPT (GPT-4o) through its visual reasoning capability.}
\label{fig:alt-example}
\end{figure}

\newpage

\footnotesize
\begin{center}
\begin{longtable}{|p{2cm}|p{3cm}|p{5cm}|p{3cm}|}
\caption{Some examples of accessibility errors, our query, corresponding suggestions from ChatGPT, and the manual interventions.} \label{table:accessibility-errors} \\ \hline
\textbf{Error Type} & \textbf{Query} & \textbf{ChatGPT Suggestions} & \textbf{Manual Adjustments / Actions} \\ \hline
\endfirsthead

Missing skip-to-content link & 
\texttt{Given the following source code <src-code>, please add skip to main content link.} & 
To add a "Skip to main content" link for better keyboard navigation and accessibility, follow these steps:
1) Add a hidden link at the top of the page that users can access via keyboard (usually the Tab key). This link will become visible when focused and allow users to skip repetitive content like the navigation bar.
2) Ensure the link moves focus directly to the main content section when activated.
Here’s how to implement it in your HTML and CSS: <HTML src-code>, <CSS src-code>& 
Add appropriate hyper reference to the link that directs to the main-landmark \\ \hline

Drop-down character focus  & 
\texttt{When the focus is on the list-box, I should be able to write any character and the list-box should open and move the visual indicator to the first option that matches the letter I wrote.} & 
To implement the functionality where the user can type any character when focused on the combobox (language dropdown), and the listbox opens with the visual indicator (highlight) moving to the first option that matches the typed character, we can enhance the existing code by adding a feature that searches through the list based on the user's input.

Steps for Implementing Search-Based Navigation: <steps>.
Here’s how we can modify the existing code to handle this:

Updated JavaScript Code with Search Functionality: <JS src-code>. & 
Replace the existing code with the improved/suggested code in the JavaScript file. If error persists, use console outputs and give feedback to ChatGPT. \\ \hline

Drop-down same character functionality & 
\texttt{The multi-character functionality is fixed,  but now the following functionality is gone again:
"If the same character is typed in succession, visual focus cycles among the options starting with that character."
I want you to add both functionalities.} & 
To achieve both functionalities — multi-character search and cycling among options when the same character is typed in succession — we can merge the two features into a cohesive solution.
Plan:
1) Multi-character search: We will track and accumulate typed characters for quick typing.
2) Same-character cycling: If the same character is typed repeatedly, the search string will be one character long, and it will cycle through options starting with that character.
3) Reset mechanism: We will reset the search string after a delay (1 second) to start fresh.
Updated Code to Support Both Features:  <JS src-code> & 
Ensure the both functionalities work, if they don't give either a feedback by telling that the issue persists and take a different approach or using debugging and providing console output.  \\ \hline

Drop-down functionalities lost, while adding the functionality for selecting last focused option upon exiting the focus & 
\texttt{Now even pressing a key doesn't work. Please revert back to the following code where both full string matching and cycling functionality were working correctly. Your task is to add the functionality where the last option should be selected if the focus moves to another element in the webpage without disturbing the other two functionalities: <JS src-code>} & 
Let's restore the working code for the two main features—cycling through options and matching multiple key inputs—and then we'll add the third feature where, if the focus moves away from the dropdown (e.g., by pressing Tab), the currently selected option is finalized, and the dropdown closes.

Here’s the updated code that includes all three functionalities: <JS src-code>

Key Features: <features explained> & 
Check the following: 1) Did ChatGPT fix the previously working functionalities? 2) Does the last focused item  get selected when exiting focus?  \\ \hline

 Pressing space/enter on navigation bar items & 
\texttt{When I press any of the menu items (except language), the focus should move to the associated section after scrolling down. 
When I navigate to the language drop-down, it should just open the drop-down and behave normally. 
Here is the entire code again: <HTML src-code> + <CSS src-code> + <JS src-code>} & 
To implement your requested behavior where:
Pressing any menu item (except for the language dropdown) scrolls to the associated section and moves focus to that section.
Pressing Enter or Space on the language dropdown should just open the dropdown and behave normally (without scrolling down the page).
We can update the code accordingly.
Here’s the complete solution: <src-code> & 
Copy the respective src-codes and replace the existing snippets (HTML, CSS, JS), check for correctness of the functionality.  \\ \hline

Carousel stop/pause movement & 
\texttt{I want you to add some sort of mechanism  to pause or stop the movement of carousel: <carousel section src-code>} & 
To add functionality for pausing and restarting the automatic movement of the carousel, we can introduce controls (like play/pause buttons) and also make the carousel stop when the user hovers over it. Here’s how you can implement this:
JavaScript to Handle Play/Pause and Hover Pausing
Add JavaScript to pause/resume the carousel automatically and to handle mouse hover to stop the movement when the user hovers over the carousel.
Add Play/Pause Buttons: This will allow users to manually control the movement.
JavaScript: <JS src-code>.
HTML: Add Play/Pause Buttons
Below the carousel, add Play and Pause buttons. You can style them to be positioned wherever you want, such as in the top-right corner of the carousel. <HTML src-code>.
CSS for Play/Pause Buttons: <CSS src-code> & 
Place the stop/pause button HTML code in the appropriate place in the carousel section. Update JS code. Check for pause functionality. Provide feedback if issue persists.  \\ \hline

Hero section focus indicator & 
\texttt{This is not good. Don't disturb the components such as the play button. Only provide a visual indicator all around the hero section.} & 
To provide a focus indicator around the hero section without disturbing the internal components (such as the play button), we can apply a simple outline on the hero section itself without using pseudo-elements or other effects that might interfere with the internal layout.
Here's how you can add a visible outline around the entire hero section: <CSS src-code> & 
Update the CSS, check the focus indicator.  \\ \hline

Pricing table accessibility (General) & 
\texttt{I want you now to do the following for the pricing table:
1) Highlight Focus: When navigating a table using the keyboard, make sure that the focused cell or row is visually highlighted. This helps keyboard users understand their position within the table.
2) Clear Header Labels: Ensure that all column and row headers are descriptive. This helps users understand the meaning of the data in each cell.
3) Focus Management: Ensure that the table can be navigated easily with a keyboard. Users should be able to tab through the cells or use arrow keys (depending on the functionality provided by assistive technologies).
4) <caption>: Use the <caption> element to provide a concise description of the table’s purpose or content. This is read by screen readers before the table content and helps users understand what the table is about.
5) Descriptive Summary: Though the <summary> attribute is deprecated in HTML5, you can provide a visually hidden description before the table for users with screen readers. This can be done using a <div> with aria-hidden="false".} & 
To achieve accessibility improvements in the pricing table, let's implement the requested features step-by-step:
Highlight Focus: Add CSS to visually highlight the focused row or cell within the table.
Clear Header Labels: Ensure the column headers are descriptive.
Focus Management: Allow keyboard navigation to make it easy for users to navigate through table cells.
Use <caption> for Table Description: Add a <caption> element to describe the table’s purpose.
Descriptive Summary for Screen Readers: Provide a visually hidden description for screen reader users.
Here’s how the revised HTML and CSS would look: <HTML CSS src-code> + <updated JS src-code>& 
Copy the respective source codes. Check general accessibility, as described in the prompt.   \\ \hline

Login/register form labels & 
\texttt{<screenshot of the login/register forms>. The labels should be on the left of the fields.} & 
To place the labels on the left of the input fields, you can apply some CSS changes to align the labels and inputs in a row layout. Here is the relevant update to your CSS and HTML code:

Updated CSS: <CSS src-code> & 
Update the CSS and give a feedback.  \\ \hline

Pressing space/enter on drop-down seems to put focus on a navigation bar item& 
\texttt{<screenshot of the opened drop-down>. When I press space or enter while on the "language" item, the "about" navigation bar item seems to have a visual indicator as can be seen in the screenshot. Also, pressing enter or space still seems to scroll down a bit} & 
It seems like when you are pressing Enter or Space on the "Language" dropdown, the About menu item is also receiving a visual focus, and pressing Enter or Space still causes some scrolling. This could be due to improper handling of focus states or overlapping styles for these elements.
Here’s how you can fix the issues:
Fixing Visual Focus Issue (with "About" menu item):
Ensure that only the Language dropdown gets the visual focus and not any other menu item when Enter or Space is pressed. This likely happens because the focus isn’t set properly when the dropdown opens.
Preventing Scrolling:
Revisit how the event handling is done for the Enter or Space key so that scrolling doesn’t occur when the dropdown opens.
Let’s fix these problems step by step.
JavaScript Fixes: <JS src-code> & 
Update the JS, ensure that only the Language drop-down gets the visual focus and not any other menu item when Enter or Space is pressed. Give feedback if the issue persists.\\ \hline

Alt text missing & 
\texttt{<screenshot of the image>. I'm going to give you images now. Describe them following the guidelines described earlier.} & 
Alt text: "Promotional poster for 'Baptiste Season 2,' featuring the main characters—a serious-looking older man with grey hair and a woman with a thoughtful expression. They stand in front of a bridge with a faint city map overlay in the background." & 
Add the text to \textit{alt} attribute.  \\ \hline

Icons in the footer not aligned properly & 
\texttt{<screenshot of the icons>. They should be inline.} & 
To display the payment method icons inline, you can use CSS to make sure they appear next to each other in a row. Here’s the updated code with inline styling for the payment methods: <HTML src-code> & 
Update the HTML code.  \\ \hline

Play button contrast & 
\texttt{<screenshot of the play button with background>. Around this background, what do you think is the most appropriate color for the outline for maximum contrast} & 
Given the background color in the image, a dark outline color would provide the best contrast around the yellow play button. Since the background is a warm, earthy tone, a color like dark brown or black would stand out well. Here are two specific options that would work well:
Black (\#000000) - This is the highest contrast and ensures maximum visibility.
Dark Brown (\#4D2C0C) - This color would create a strong outline that’s distinct without being as stark as black, yet still provides contrast.
If you’re aiming for maximum accessibility and visibility, black is the optimal choice, as it provides the highest contrast against the yellow button and the background.
Here’s how you could apply it: <CSS src-code> & 
Update the CSS, give further screenshots based feedback if the contrast error persists. \\ \hline

Table headers contrast & 
\texttt{<screenshot of the table>. The table header text for columns does not have a good contrast} & 
To improve the contrast of the table header text against the yellow background, you can update the color of the header text to a darker shade that enhances readability. Here’s how you can adjust it:
CSS Code Update:
table th {
    background-color: \#ffd700; /* Existing yellow background */
    color: \#2c2c2c; /* Dark gray text for better contrast */
    padding: 12px; /* Optional: Adjust padding for readability */
    font-weight: bold;
    text-align: left;
}
Explanation
The \#2c2c2c color is a dark gray that provides better contrast with the yellow background than white or light gray, ensuring the headers remain easily readable while maintaining visual hierarchy. Adjustments to padding can also enhance readability without changing the overall design. & 
Update the CSS, give further screenshots based feedback if the contrast error persists.  \\ \hline

\end{longtable}
\end{center}



\begin{figure*}
\centering
\includegraphics[width=0.75\textwidth]{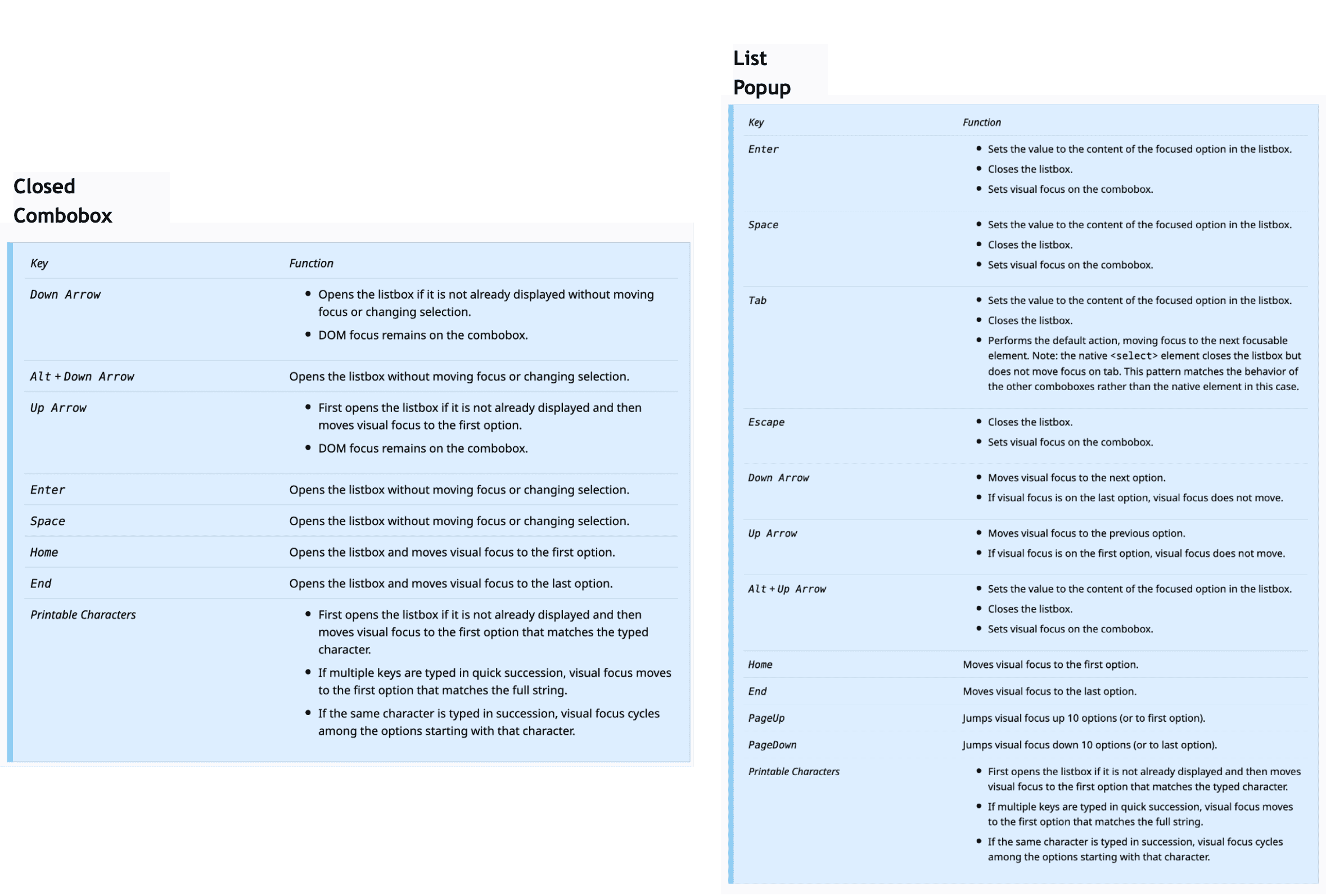}
\caption{Rules of accessibility for a dropdown \cite{w3c_wcag_2024}}
\label{fig:dropdown}
\end{figure*}

\begin{figure*}
\centering
\includegraphics[width=0.75\textwidth]{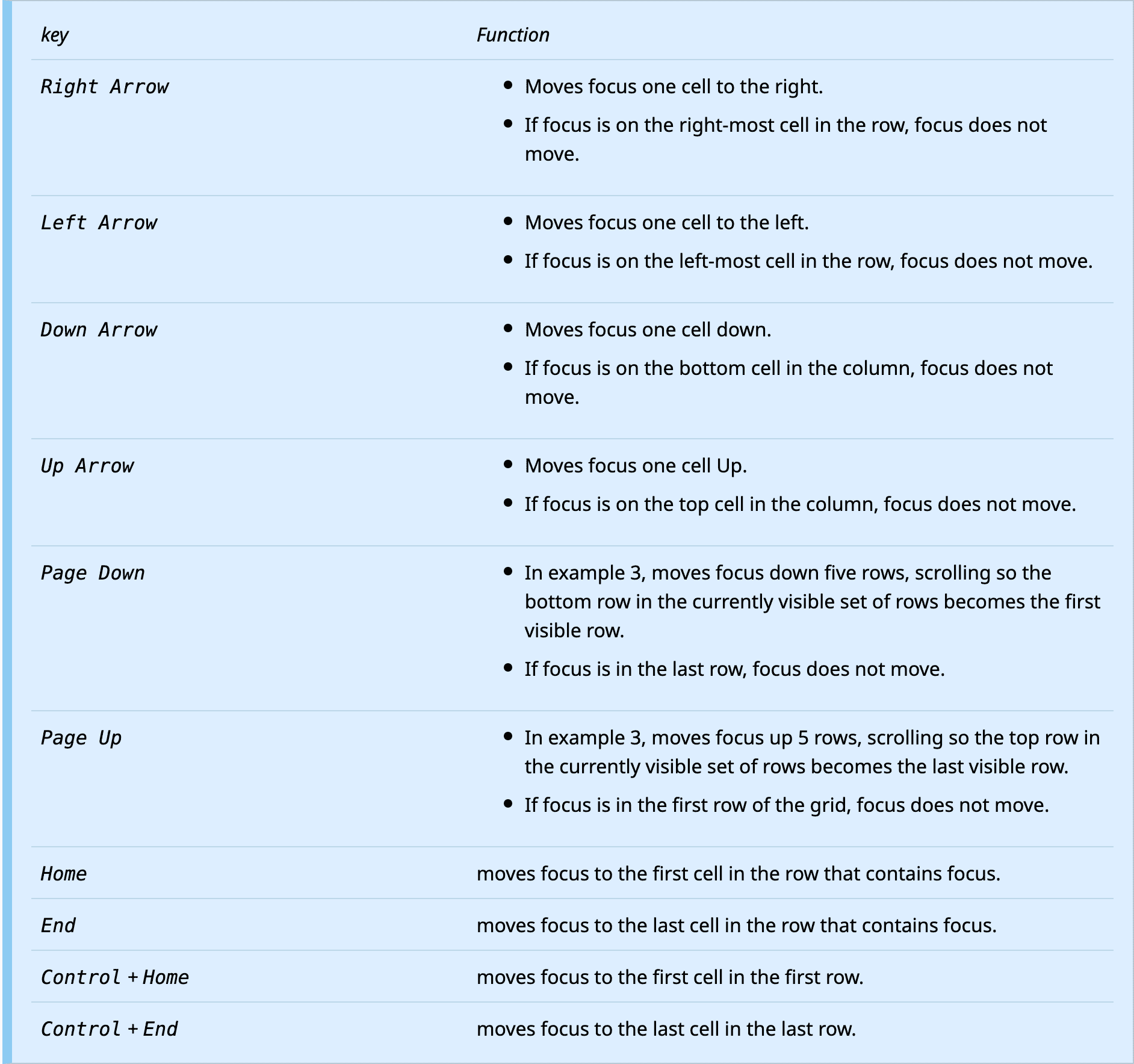}
\caption{Rules of accessibility for a table \cite{w3c_wcag_2024}}
\label{fig:table}
\end{figure*}


\end{document}